\documentclass[epj]{svjour}
\usepackage{graphicx}
\begin{document}
\title{Arches and contact forces in a granular pile}
\author{C. Manuel Carlevaro\inst{1,2} \and Luis A. Pugnaloni\inst{1}% etc
}                     % Do not remove
\institute{Instituto de F\'{\i}sica de L\'{\i}quidos y Sistemas Biol\'{o}gicos (CONICET La Plata, UNLP), Calle 59 Nro 789, 1900 La Plata, Argentina \and Universidad Tecnol\'ogica Nacional - FRBA, UDB F\'{\i}sica, Mozart 2300, C1407IVT Buenos Aires, Argentina.}
\date{Received: date / Revised version: date}
\abstract{
Assemblies of granular particles mechanically stable under their own weight contain arches. These are structural units identified as sets of mutually stable grains. It is generally assumed that these arches shield the weight above them and should bear most of the stress in the system. We test such hypothesis by studying the stress born by in-arch and out-of-arch grains. We show that, indeed, particles in arches withstand larger stresses. In particular, the isotropic stress tends to be larger for in-arch-grains whereas the anisotropic component is marginally distinguishable between the two types of particles. The contact force distributions demonstrate that an exponential tail (compatible with the maximization of entropy under no extra constraints) is followed only by the out-of-arch contacts. In-arch contacts seem to be compatible with a Gaussian distribution consistent with a recently introduced approach that takes into account constraints imposed by the local force balance on grains.
%
%\PACS{
%      {PACS-key}{discribing text of that key}   \and
%      {PACS-key}{discribing text of that key}
%     } % end of PACS codes
} %end of abstract
\maketitle
\section{Introduction}

The study of mechanically stable granular beds has become a major point of interest in granular Physics. Studies range from analysis of force network properties \cite{ostojic,behringer,mueth,peters} to structural characterization \cite{latzel,aste,torquato} to thermodynamic and statistical descriptions \cite{edwards1,henkes,tighe,snoeijer1,pugnaloni1}. A recurrent question in the subject is related to the existence or not of a relation between force chains and arches \cite{mehta1,mehta3}. 

Arches (or bridges) are multiparticle structures where all grains are mutually stable \cite{mehta1,pugnaloni2,pugnaloni3,jenkins}, i.e., fixing the positions of all other particles in the assembly, the removal of any particle in the arch leads to the destabilization of the other particles in it. For an arch to be formed, it is necessary (although not sufficient) that two or more falling particles be in contact at the time they reach mechanical equilibrium in order to create mutually stabilizing structures \cite{arevalo}. Like arches in architecture, granular arches are assumed to sustain the weight of the material above. 

Highly stressed grains in static deposits are generally found to form linear structures: the so-called force chains. Notice that, the term ``arch'' is sometimes used \cite{lovoll,dorbolo,nicodemi} to refer to these force chains and not to the mutually stabilizing structures defined above. Likewise, the term ``dynamic arch'' has been used to refer to ephemeral structures that choke the flow of grains \cite{luding1}. We have to distinguish between these usages and the classical meaning we adhere to in this work: an arch is a mechanically stable structure of mutually stabilizing bodies.

Force chains are a clear spatial heterogeneity of the contact force network. The distribution of contact forces (both normal and tangential) shows no bimodal character. However, the spatial distribution of large and small forces is heterogeneous with large forces developing a somewhat open stringy network that encloses regions of grains that sustain little weight \cite{radjai}. 

To what extent the bimodal spatial distribution of forces is related to the mutually stable structures (arches) has not been assessed so far. A correlation as been pointed out \cite{mehta1,pugnaloni2,mehta2} between the distribution of horizontal span of arches in a granular pile and the distribution of normal forces at the grain contacts. Therefore, it is assumed that a strong correlation has to be present between arches and highly stressed grains in a granular deposit. In this paper, we assess this general belief in the frame of a simulation of granular packings prepared by tapping. The results provide additional information on the validity of basic assumptions made in the statistical description of granular packings.

\section{Simulation model}

To simulate packings of gains we have followed the standard techniques on discrete element methods (see for example Refs. \cite{cundall,poschel,schafer}). We used a velocity Verlet algorithm to integrate the Newton equations for $N$ monosized disks (diameter $d$) in a rectangular box of width $L$. We studied two system sizes: (i) $N=512$ with $L=12.39 d$ and (ii) $N=2048$ with $L=24.78 d$. The non-commensurate box is chosen to prevent crystallization to some extent. The larger system is roughly twice as tall as the smaller system. However, this depends on the actual packing fraction obtained for a given preparation protocol. 

The disk--disk and disk--wall contact interaction comprises a linear spring--dashpot in the normal direction 
\begin{equation}
F_{\rm{n}}=k_{\rm{n}}\xi -\gamma _{\rm{n}}v_{i,j}^{\rm{n}} \label{eqmodel1}
\end{equation} 
and a tangential friction force 
\begin{equation}
F_{\rm{t}}=-\min \left( \mu |F_{\rm{n}}|,|F_{\rm{s}}|\right) \cdot \rm{sign}\left( \zeta \right) \label{eqmodel2}
\end{equation} 
that implements the Coulomb criterion to switch between dynamic and static friction \cite{arevalo,pugnaloni4}. 

In Eqs. (\ref{eqmodel1})--(\ref{eqmodel2}), $\xi =d-\left\vert \mathbf{r}_{ij}\right\vert $ is the particle--particle overlap, $\mathbf{r}_{ij}$ represents the center-to-center vector between particles $i$ and $j$, $v_{i,j}^{\rm{n}}$ is the relative normal velocity, $F_{\rm{s}}=-k_{\rm{s}}\zeta -\gamma _{\rm{s}}v_{i,j}^{\rm{t}}$ is the static friction force, $\zeta \left( t\right) =\int_{t_{0}}^{t}v_{i,j}^{\rm{t}}\left( t^{\prime}\right) dt^{\prime }$ is the relative shear displacement, $v_{i,j}^{\rm{t}}=\dot{\mathbf{r}}_{ij}\cdot \mathbf{s}+\frac{1}{2}d\left(\omega _{i}+\omega _{j}\right)$ is the relative tangential velocity, and $\mathbf{s}$ is a unit vector normal to $\mathbf{r}_{ij}$. The shear displacement $\zeta $ is calculated by integrating $v_{i,j}^{\rm{t}}$ from the beginning of the contact (i.e., $t=t_{0}$). The disk--wall interaction is calculated considering the wall as an infinite radius, infinite mass disk. Other than these, the interaction parameters are the same as for the disk-disk interaction.

In these simulation we used the following set of parameters: dynamic friction coefficient $\mu = 0.5$, normal spring stiffness $k_n = 10^5 (mg/d)$, normal viscous damping $\gamma_n = 300 (m\sqrt{g/d})$, tangential spring stiffness $k_s= \frac{2}{7}k_n$, and tangential viscous damping $\gamma_s = 200 (m\sqrt{g/d})$. The integration time step is $\delta = 10^{-4} \sqrt{d/g}$. Units are reduced with the diameter of the disks, $d$, the disk mass, $m$, and the acceleration of gravity, $g$.

In order to investigate reproducible states, we implement a tapping protocol. The system is initially deposited from a dilute configuration in which particles have no contacts nor overlaps. After the grains reach mechanical equilibrium, the system is tapped with a given amplitude and left to come back to mechanical equilibrium. After many taps of given amplitude, the system reaches a steady state where the properties of the static configurations generated have well defined mean values and fluctuations. The steady state properties are independent of the initial conditions and are reproducible \cite{pugnaloni1,ribiere,pugnaloni5}. We generate a number of static packings after the steady state is reached to average quantities. Different steady states are generated by changing the tap amplitude.
 
Tapping is simulated by moving the confining box in the vertical direction following a half sine wave trajectory of frequency $\nu = \pi /2 (g/d)^{1/2}$. The intensity of the excitation is controlled through the amplitude, $A$, of the sinusoidal trajectory; and it is characterized by the parameter $\Gamma = A (2\pi\nu)^2/g$. A new tap is applied only after the system has come to mechanical equilibrium, which is defined via the stability of each particle-particle contact \cite{arevalo}. Averages were taken over 500 taps (configurations) in the steady state corresponding to each value of $\Gamma$ after the $500$ initial taps were discarded to avoid any transient.

\section{Identification of arches}

\begin{figure}[b]
\centering
\includegraphics[width=0.9\columnwidth]{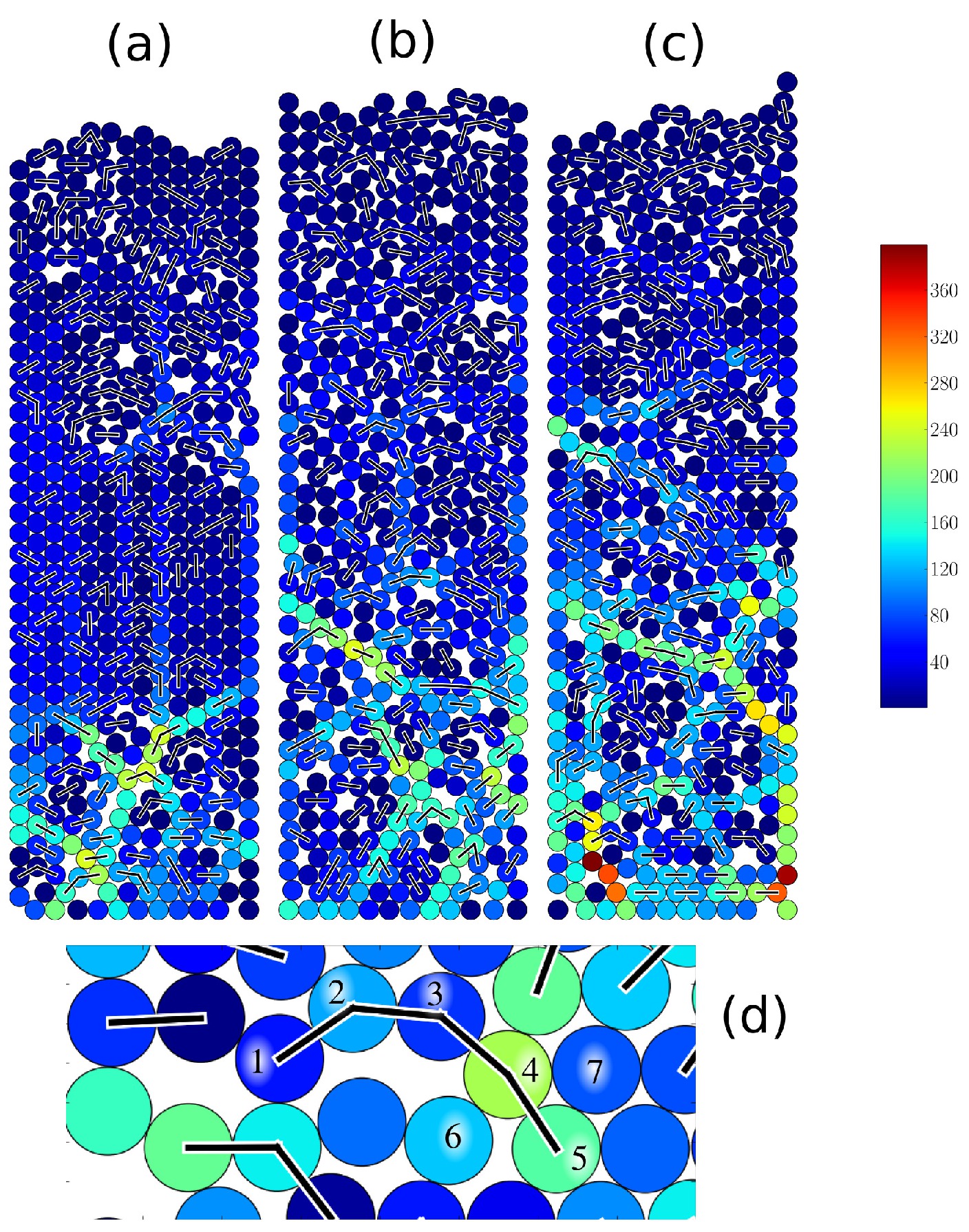}
\caption {(Color online). Sample images of the simulated granular columns ($N=512$) for different $\Gamma$: (a) $\Gamma = 2.19$, (b) $\Gamma = 4.93$ and (c) $\Gamma = 15.39$. The color code indicates the trace, Tr$(\sigma)$, of the stress tensor for each particle in units of $mg/d$. The joining segments indicate the arches detected in the system. (d) A closeup on a 5-particle arch. See main text for a description on the supporting contacts of particle 4.}
\label{fig1}
\end{figure}

Details on the algorithms used to identify arches can be found in previous works \cite{pugnaloni2,pugnaloni3,arevalo}. Briefly, we need first to identify the supporting grains of each particle in the packing. In 2D, there are two disks that support any given grain. Two grains in contact with a given particle are able to provide support if the segment defined by the contact points lies below the center of mass of the particle. Some of these supporting contacts may be provided by the walls of the container. Then, we find all mutually stable particles. Two grains A and B are mutually stable if A supports B and B supports A. Arches are defined as sets of particles connected through these mutually stabilizing contacts (MSC). 

The fact that the supporting particles of each grain have to be known implies that contacts, and the chronological order in which they form, have to be clearly defined in the model. Figure \ref{fig1} shows some examples of the packings generated where arches are indicated by joining segments. In Fig. \ref{fig1}(d), a close view of an arch detected in a given packing (formed by particles 1 to 5) is displayed. Particle 4 is an example of a grain whose pair of stabilizing particles is ambiguous form the limited information provided by the snapshot; discrimination requires chronological information. The pairs 3-5, 5-6 and 6-7 comply with the condition that the center of mass of grain 4 is above the segment that joins the corresponding contact points. However, the contacts with grains 3 and 5 where formed first and for that reason these particles are considered to be the supporting pair of grain 4. To identify the contacts that support each particle we use an algorithm that has been previously designed to work with molecular dynamic type simulations \cite{arevalo}.

\section{Single particle stress tensor}

We measure the stress tensor $\sigma_i$ for grain $i$ as \cite{latzel}:

\begin{equation}
 \sigma_i^{\alpha\beta}= \frac{1}{\pi (d/2)^2} \sum_{j=1}^{N_c}{f_{ij}^\alpha b_{ij}^\beta}, \label{eq1}
\end{equation}
where, $f_{ij}$ is the force exerted by grain $j$ on grain $i$ and $b_{ij}$ is the branch vector that goes from the center of grain $i$ to the contact point with grain $j$. The sum runs over the $N_c$ particles in contact with particle $i$.  

The pressure (or isotropic stress) is given by the trace, $\rm{Tr}(\sigma)$, of $\sigma$ whereas the anisotropic component is characterized by the deviatoric stress $s$

\begin{equation}
 s_i^{\alpha\beta}=\sigma_i^{\alpha\beta}-\frac{\delta_{\alpha\beta}}{3}\sum_{\gamma}{\sigma_i^{\gamma\gamma}} \label{eq2}
\end{equation}

We use $\rm{Dev}(\sigma)=\sigma^{zz}-\sigma^{xx}$ to characterize the anisotropic component. In average, our packings under gravity present $\sigma^{xx} < \sigma^{zz}$ (i.e., the vertical component is higher than the horizontal component).

The principal directions of the stress vary form configuration to configuration during tapping. However, these fluctuations are very small since the shear component $\sigma^{xz}$ is less than 1\% of $\rm{Tr}(\sigma)$ in all our packings.

\begin{figure}[t]
\centering
\includegraphics[width=1\columnwidth]{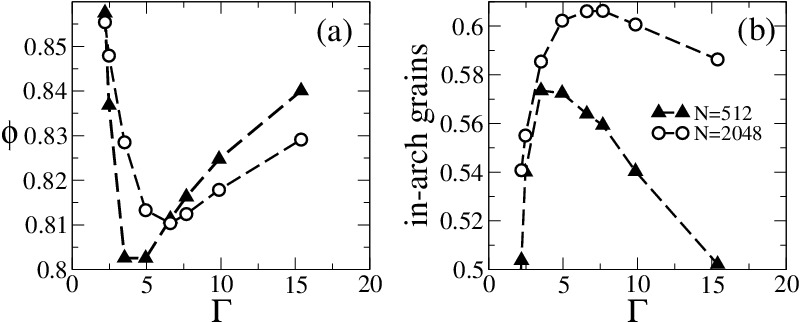}
\includegraphics[width=0.6\columnwidth]{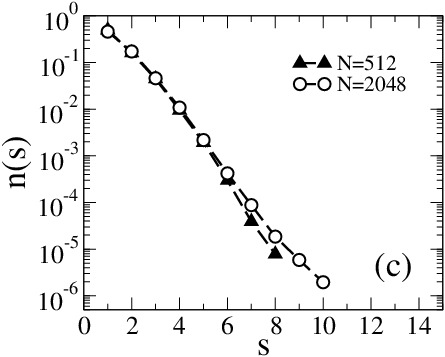}
\caption {(a) The mean packing fraction, $\phi$, as a function of tap intensity $\Gamma$. (b) Fraction of in-arch grains as a function of $\Gamma$. (c) The arch size distribution $n(s)$ for $\Gamma=2.19$. Results obtained for the two system sizes: $N=512$ (solid triangles) and $N=2048$ (open circles). }
\label{fig2}
\end{figure}

\section{General properties of the deposits}

In Fig. \ref{fig2}(a), we report the mean packing fraction, $\phi$, as a function of the tap intensity, $\Gamma$. The packing fraction was measured in a slab of the bed that covers 50\% of the height of the column and is vertically centered with the center of mass of the system. The values of $\phi$ for the smaller system is affected by the presence of the lateral walls, which tend to reduce the packing fraction. $\phi$ presents a minimum at intermediate $\Gamma$ as previously observed in various models \cite{pugnaloni4,gago} and experiments \cite{pugnaloni1,pugnaloni5}. This minimum in $\phi$ is related with the existence of a maximum in the number of grains involved in arches [see Fig. \ref{fig2}(b)]. A description of the mechanisms that lead to the existence of a maximum in the number of grains involved in arches can be found in Ref. \cite{pugnaloni4}. 

In spite of the system being monodisperse, the packings obtained present only partial crystallization. This is due to the non-commensurate simulation box. Even if very ordered packings were obtained, the contact forces (the main focus of this paper) have been found to display similar statistics to the one shown by disordered packings \cite{blair}. We have also assessed the structural anisotropy through the fabric tensor. We found that all our packings present a deviatoric fabric of less than 5\% of the fabric trace. Therefore, the structural anisotropy is rather small.

The distribution, $n(s)$, of the sizes of the arches found in the packings is shown in Fig. \ref{fig2}(c). Here, $n(s)$ is the fraction of arches consisting in $s$ grains, with $n(s=1)$ the fraction of grains that do not belong to any arch. As we can see, $n(s)$ is not affected by the system boundaries and arches of more than 10 disks have not been detected even in the 24-disk-wide system (i.e., $N=2048$).

Figure \ref{fig1} shows some examples of the distribution of pressures and arches inside a granular pile. As it is to be expected, particles are subjected to higher pressures, in average, in the lower part of the pile as compared with the upper layers. The system does not display force chains that span the system from top to bottom as is commonly seen in many experiments and simulations. This is due to the fact that the system is in mechanical equilibrium under its own weight; no external compression is applied to the sample in any direction.

\begin{figure}[b]
\centering
\includegraphics[width=0.6\columnwidth]{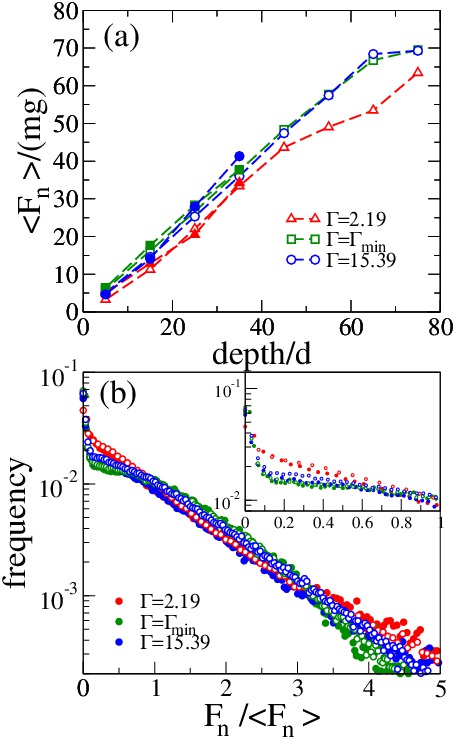}
\caption {(Color online). (a) Mean normal contact force $\langle F_n \rangle$ as a function of the depth into the granular column.  (b) PDF of the normal contact force. We consider two system sizes: $N=512$ (solid symbols) and $N=2048$ (open symbols). Results for three different tap intensities are reported (see legend). The intermediate value corresponds to the value of $\Gamma$ that yields the minimum $\phi$ for the given system size (i.e., $\Gamma=4.93$ for $N=512$ and $\Gamma=6.59$ for $N=2048$). The inset in part (b) is a close up for forces below the mean.}
\label{fig3}
\end{figure}

In Fig. \ref{fig3}, we show the normal component of the contact forces for three different tap intensities (the lower and the higher $\Gamma$ studied, and the value $\Gamma_{\rm{min}}$ that leads to the minimum $\phi$ for the given system size). The mean normal contact force $\langle F_{\rm{n}} \rangle$ increases rather linearly with the depth into the packing and is little dependent on the system size for any given depth [see Fig. \ref{fig3}(a)]. Only the system with 2048 grains display a hint of Janssen saturation in the deeper layers. Small differences in $\langle F_{\rm{n}} \rangle$ can be observed between packings obtained with different $\Gamma$. In particular, for the lowest tap intensity considered,  $\langle F_{\rm{n}} \rangle$ is smaller at all depths.

Figure \ref{fig3}(b) presents the normal contact force distribution for a depth of $35 d$ (this corresponds to the lower part of the smaller system and to the middle section of the larger system). All grain-grain contacts that lay within a slab $10 d$-wide centered at a depth $35 d$ are considered. Taking narrower slabs leads to similar results. As we can see, the PDF of $F_{\rm{n}}$ coincides for both system sizes. We have seen that the tangential contact forces also show consistent results when systems of different sizes are compared by looking into slabs at the same depth. 

There exist a current debate on whether the tail of these distributions are or not exponentials \cite{tighe,eerd}. Exponential tails for forces above the mean contact force have been reported by a number of authors considering granular packs subjected to external compression \cite{behringer,mueth,snoeijer2} or stable under their own weight \cite{lovoll}. As we can see in Fig. \ref{fig3}(b), for low $\Gamma$, we observe a clear exponential tail. However, a faster than exponential decay seem to be followed by the rest of the packings. Tighe et al. \cite{tighe} have argued that some reported exponential tails are perhaps Gaussians. The behavior for very small forces [see inset to Fig. \ref{fig3}(b)] resembles the weak divergence found for packings without external compression when bulk contacts (as opposed to contacts made between the grains and the container) are considered \cite{snoeijer2}.

In order to compare results from different system sizes, the remaining of the paper, unless otherwise specified, will refer to measurements made in a slab $10 d$-wide centered at a depth $35 d$.

\section{Contact forces and arches}

\begin{figure}[b]
\centering
\includegraphics[width=0.6\columnwidth]{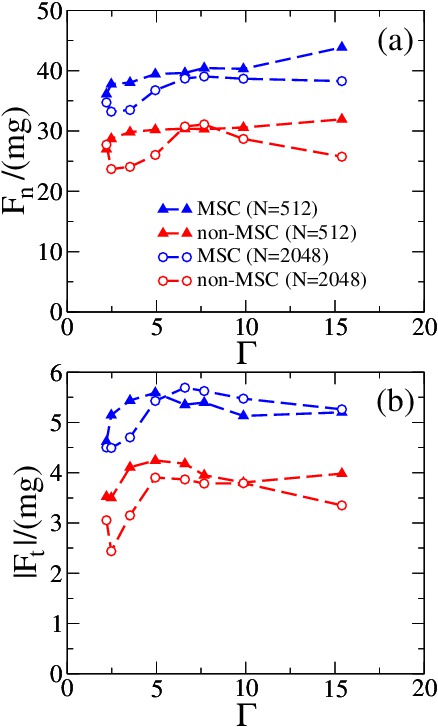}
\caption {(Color online). The mean value of the contact force for mutually stabilizing contacts (blue) and non-mutually stabilizing contacts (red). (a) Normal contact forces, and (b) tangential contact forces. Results obtained for the two system sizes: $N=512$ (solid triangles) and $N=2048$ (open circles).}.
\label{fig4}
\end{figure}

It can be observed from Fig. \ref{fig1} that, at any depth into the pile, grains can present high and low stress irrespective of whether they belong to an arch or not. Also, force chains do not coincide with arches although arches form part of portions of these chains. For a more quantitative analysis we plot in Fig. \ref{fig4} the mean value of the contact forces (normal and tangential to the contact in a slab at $35 d$ of depth) as a function of $\Gamma$. MSC (mutually stabilizing contacts) and non-MSC have been separated in the analysis. Although some small differences are observed between the results for the two system sizes studied, the general trends are quite similar. It is clear that MSC (i.e., contacts within arches) have, in average, larger (roughly 50\%) normal and tangential forces. This supports the idea that arches bear most of the stress in the system and that force chains and arches must be correlated. 

\begin{figure}[t]
\centering
\includegraphics[width=0.6\columnwidth]{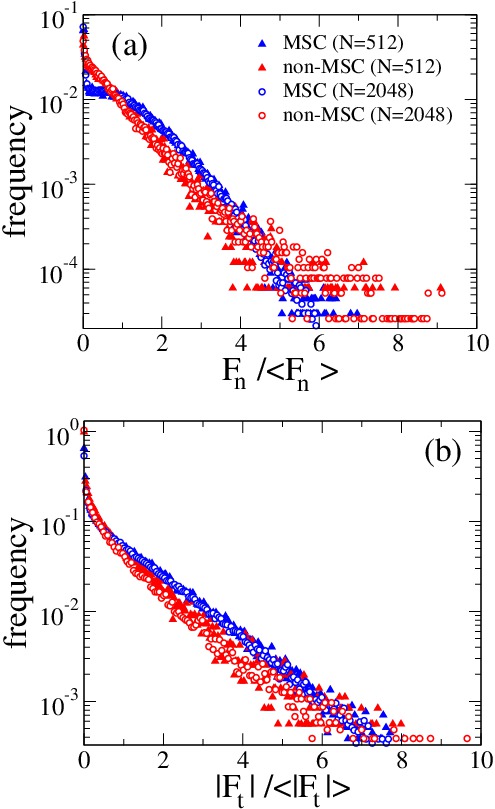}
\caption {(Color online). The PDF of contact forces for MSC (blue) and non-MSC (red) for $\Gamma=\Gamma_{min}$. (a) Normal contact forces, and (b) tangential contact forces. Results obtained for the two system sizes: $N=512$ (solid triangles) and $N=2048$ (open circles).}
\label{fig5}
\end{figure}

Figure \ref{fig5}(a) shows that the distribution of contact forces for MSC and non-MSC are clearly distinguishable for the normal component. Although, we present the distribution obtained for $\Gamma_{\rm{min}}$, most packings display the same general features (some exceptions regarding packings prepared at low $\Gamma$ are discussed below). The mild divergence for very small forces is still present for both distributions. Despite the difference, there is not a clear separation of the two populations of contacts. The bimodal character observed in the spatial distribution of contacts seems to be poorly correlated with MSC.

As we can see from Fig. \ref{fig5}(a), the PDF for non-MSC present a clear exponential tail, whereas MSC present a faster decay. The non-MSC PDF corresponds to an exponential decay even for forces below the mean. The exponential PDF has a well established statistical explanation. If the mean force is set and all contact force states are equally probable, an exponential distribution of contact forces maximizes the entropy (defined as the logarithm of the number of contact force states) \cite{edwards2,evesque}. 

MSC seem to have a distribution compatible with a Gaussian tail, or at least a faster than exponential tail. It seems that the deviation form an exponential in the full PDFs reported in the literature [and in Fig. \ref{fig3}(b)] seem to be due to the presence of MSC (and therefore the presence of arches). The immediate conclusion is that the presence of arches prevents us from making some of the basic assumptions on the contact forces to render a simplified statistical analysis. In particular, arches introduce force balance constraints that need to be accounted for. Tighe et al. \cite{tighe} have shown that force balance constraints (the conservation of the total area of the Maxwell reciprocal tiling) can be introduced in a force ensemble. These have led to Gaussian contact force distributions. Notice however that this theoretical approach yields the same Gaussian distribution irrespective of the existence of arches in the packing.

Figure \ref{fig5}(b) shows that the tangential components of the contact forces have a much subtle difference between the distribution for MSC and non-MSC. Again, MSC present a somewhat faster-than-exponential tail in contrast with the non-MSC.

We now focus in the results for the smallest tap intensity reported. As we mentioned, Fig. \ref{fig3}(b) shows that for $\Gamma=2.19$ the PDF for normal contacts presents a clear exponential tail, in contrast with the packings generated with stronger taps. Separating MSC and non-MSC in the analysis leads to two exponential tails (presenting slightly different slopes). We speculated that there could be fewer MSC in these packings than in packings obtained with stronger taps. However, these packings present similar number of MSC as compared with packings that show a faster-than-exponential tail. The main difference we have been able to find is that these packings have, in comparison, fewer arches composed of three or more grains. It seems that arches composed of three or more particles are the responsible for introducing strong force balance constraints that render the PDF non-exponential. Some reports of pure exponential decays can be found in previous studies. Blair et al. mentioned that a pure exponential was found in some cases depending on the history of the packing \cite{blair}. Makse et al. found pure exponentials too in simulations of isotropically compressed grains, which may develop structures without arches \cite{makse}. We believe the preparation history of these packings may have lead to a small presence of arches composed of three or more grains.

\section{Stress tensor and arches}  

\begin{figure}[t]
\centering
\includegraphics[width=0.6\columnwidth]{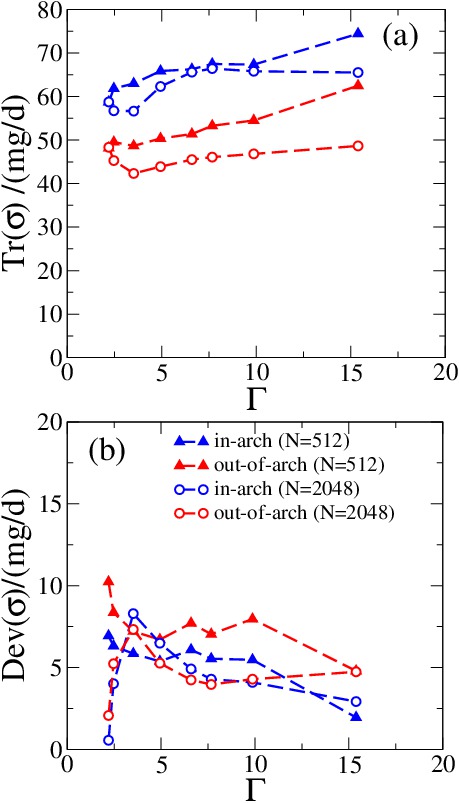}
\caption {(Color online). Mean stress tensor for in-arch (blue) and out-of-arch (red) grains. (a) The trace $\rm{Tr}(\sigma)$ of the stress, and (b) the deviator $\rm{Dev}(\sigma)$. Results obtained for the two system sizes: $N=512$ (solid triangles) and $N=2048$ (open circles).}
\label{fig6}
\end{figure}

\begin{figure}[t]
\centering
\includegraphics[width=0.6\columnwidth]{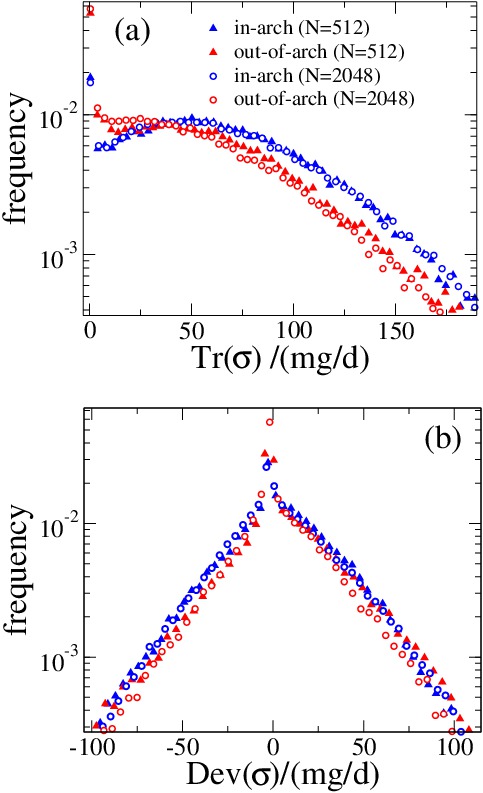}
\caption {(Color online). The PDF of the stress tensor for in-arch (blue) and out-of-arch (red) for $\Gamma=\Gamma_{min}$. (a) The trace, $\rm{Tr}(\sigma)$, of the stress tensor, and (b) the deviator $\rm{Dev}(\sigma)$. Results obtained for the two system sizes: $N=512$ (solid triangles) and $N=2048$ (open circles).}
\label{fig7}
\end{figure}

In Fig. \ref{fig6}, we show the results of an analysis similar to the previous section but now the stress tensor on each particle, as defined in Eq. (\ref{eq1}), is considered. The stress tensor accounts for both MSC and non-MSC on each grain. We separate in-arch grains from out-of-arch grains in the analysis. In-arch grains support, in average, isotropic pressures [see $\rm{Tr}(\sigma)$ in Fig. \ref{fig6}(a)] about 20\% higher than out-of-arch grains. In contrast, the anisotropic component of the stress, $\rm{Dev}(\sigma)$, seems to be rather similar for both types of grains. This implies that the actual difference between the stress tensor of in-arch and out-of-arch grains corresponds to the addition of a constant to the diagonal components (in contrast to an increse given by a multiplicative constant). We have seen that the shear stress $\sigma_{xz}$ is the same for both types of grains.

Figure \ref{fig7}(a) shows that the distribution of the isotropic stress is markedly different for in-arch and out-of-arch grains. In-arch grains present a clear maximum in the PDF of $\rm{Tr}(\sigma)$ at around the mean. Although there is not a strong separation, the maximum in the PDFs suggest that the well known bimodal character of the force network is driven by the presence of arches to some extent.

The distribution of the stress deviator is presented in Fig. \ref{fig7}(b). As we can see, the PDFs for in-arch and out-of-arch grains are almost identical. The negative values are due to the fact that some grains have $\sigma_{xx} > \sigma_{yy}$. However, in average $\sigma_{xx} < \sigma_{yy}$ and the mean deviator as defined above is always positive in our packings.

\section{Conclusions}

We have shown that MSC, which define arches, present higher normal and tangential components of the contact forces as compared with non-MSC. Grains belonging to arches are generally subjected to larger isotropic stresses but similar anisotropic stress. Therefore, particles in arches are, to some extent, different from particles that do not form arches when their contact forces are considered. This is in line with the common assumption that arches carry most of the weight in a granular deposit. 

The PDF of normal contact forces show that non-MSC follow an exponential decay whereas the MSC present a faster-than-exponential fall. This has strong implications for the basic statistical models of force distribution. In particular, it seems that MSC are the main cause for the constraints in force balance not considered in simplistic approaches. These constraints lead to the deviation of the overall-contacts PDF from the expected exponential. Indeed, packings containing a low number of large arches (arches of three or more grains) seem to fit better the exponential law.

The bimodal spatial distribution of stresses seems to be related to some extent with the presence of arches. Particles in arches present a clear maximum around the mean stress in the PDF of isotropic stress. 

It is worth mentioning that despite the correlations found between arches and force chains, there is not a one-to-one correspondence. Arches that sustain little weight can always exist in the structure since they are shielded by other arches above. This leads to the preponderance of very small forces in the distributions for MSC. Also, force chains can develop without the need of arches. A deposit carefully built by sequential deposition of grains contains no arches in the structure, yet it will present force chains.

\section*{Acknowledgements}

LAP thanks fruitful discussions with Gary C. Barker and Anita Mehta. This work was supported by CONICET (Argentina). We thank an anonymous reviewer for a insightful suggestion on the first version of the manuscript.

%
% BibTeX users please use
% \bibliographystyle{}
% \bibliography{}

\end{document}